\documentstyle[preprint,aps,epsfig,axodraw,floats]{revtex}

\begin{document}

\newcommand{\y }{\'{\i}}
\newcommand{\be}{\begin{equation}}
\newcommand{\ee}{\end{equation}}
\newcommand{\eeq}{\end{equation}}
\newcommand{\seta}{\rightarrow}
\newcommand{\beqn}{\begin{eqnarray}}
\newcommand{\eeqn}{\end{eqnarray}}
\newcommand{\bef}{\begin{figure}[hbt]}
\newcommand{\ef}{\end{figure}}
\newcommand \sig {\Sigma}
\newcommand \pt {p_T\!\!\!\!\!\!/\hskip 0.2cm}
\newcommand \rp {R_P}
\def\lsim{\raise0.3ex\hbox{$\;<$\kern-0.75em\raise-1.1ex\hbox{$\sim\;$}}}
\def\gsim{\raise0.3ex\hbox{$\;>$\kern-0.75em\raise-1.1ex\hbox{$\sim\;$}}}
\def\stau{\tilde{\tau}}
\def\stau1{\tilde{\tau}_1}
\def\stop{\tilde{t}_1}
\def\sbottom{\tilde{b}}
\def\schi{\tilde{\chi}^0_1}
\newcommand{ \slashchar }[1]{\setbox0=\hbox{$#1$}   
   \dimen0=\wd0                                     
   \setbox1=\hbox{/} \dimen1=\wd1                   
   \ifdim\dimen0>\dimen1                            
      \rlap{\hbox to \dimen0{\hfil/\hfil}}          
      #1                                            
   \else                                            
      \rlap{\hbox to \dimen1{\hfil$#1$\hfil}}       
      /                                             
   \fi}
\def\rps{\slashchar{R}_{p}}
\def\met{\slashchar{E}_{T}}
\def\fbi{\rm fb^{-1}}
\def\gev{\rm GeV}

\draft

\preprint{
\hfill$\vcenter{ 
 \hbox{\bf MADPH-99-1146} 
{\hbox{\bf hep-ph/9911442} }
}$ 
}

\title{Top-quark decay via $R$-parity violating interactions\\  
at the Tevatron}

\author{T. Han and M. B. Magro}

\address{Department of Physics, University of Wisconsin,
Madison, WI 53706, USA}

\date{January, 2000}

\vskip -0.75cm 

\maketitle

\begin{abstract}
  
\vskip-5ex 

We consider the top-quark decay
$t\to \tilde \tau b$ and $t\to \tau b \tilde{\chi}^0_1$ via explicit 
$R$-parity violating interactions in SUSY theories. 
We discuss the observability of those channels at the
Fermilab Tevatron collider. The existing Tevatron data 
indicate a 95\% confidence level upper bound on the coupling 
to be 0.94 (0.63) for a long-lived (short-lived) 
$\schi$ with $m_{\tilde\tau}=70$ GeV. 
At Tevatron Run II with an integrated luminosity of 2 (10) 
fb$^{-1}$, one can obtain 
a 2$\sigma$ constraint as 0.38 (0.24) for a long-lived $\schi$ 
and 0.29 (0.19) for a short-lived $\schi$,
beyond the current indirect limit.
\end{abstract}


\section{Introduction}

Physics associated with the top quark is fascinating. Being the heaviest
elementary particle observed so far with a mass ${m_t\approx 175}$ GeV, 
the top quark has the greatest potential to probe new physics beyond
the Standard Model (SM). Examples considered include the radiative
electroweak symmetry breaking due to the large top-quark Yukawa coupling
in supersymmetric (SUSY) grand unified theories (GUT),
and the dynamical electroweak symmetry breaking in the top-quark 
sector. Furthermore, the fact that the top quark 
is about 35 times more massive than its SU(2)-doublet partner, 
the $b$ quark, makes it an attractive
candidate for studying flavor physics.

The Fermilab Tevatron will start a new run with significant
upgrade on luminosity, reaching approximately 2 $\fbi$ in
a year. This would yield more than ten thousand $t\bar t$
pairs and thus provide a great opportunity to explore new physics
in the top-quark sector. Observing the top-quark rare decays 
may reveal information on new physics.
For instance, if there are charged scalars
such as the charged Higgs bosons ($H^\pm$), the decay $t\to H^+b$
may occur if kinematically accessible. This search has been
carried out with the existing data from the Tevatron by the CDF and D0
collaborations \cite{cdf,d0}. The non-observation of the 
predicted signal 
has established limits on the charged Higgs boson mass and its 
coupling to the top quark. In SUSY GUT models, 
one may be motivated to search for decays to scalar quarks
and gauginos  \cite{topstop}
\begin{equation}
t\to \stop \schi\quad {\rm and}\quad \sbottom \tilde \chi^+
\label{tsusy}
\end{equation}
where $\stop$ is the lighter top-squark and
$\schi$ the lightest neutralino, assumed to be
the lightest supersymmetric particle.
Furthermore, if $R$-parity ($\rp$) is broken 
\cite{rbreaking,rplimit}, then $\schi$ will not 
be stable and even possibly decay within the detector, 
leading to a different signature
in the Tevatron collider 
experiments \cite{roger,tev}.

Process Eq.~(\ref{tsusy}) may be disfavored by kinematics
for heavy SUSY particles. Also, generic $\rp$-breaking
($\rps$) interactions 
are subject to stringent experimental
constraints due to the absence of proton decay and large
flavor-changing neutral currents in the lepton and quark
sectors \cite{rplimit}. However, the bounds on operators
involving the third generation fermions are particularly weak, 
which reflects our lack of knowledge for flavor physics
in the third generation. If there exist sizeable $\rps$ 
interactions involving the third generation, then the top 
quark may directly decay to other interesting channels.
In fact, decays of the top quark and the top squark
have been studied in scenarios where $\rp$ is
spontaneously broken \cite{spon}.
In this Letter, we explore the explicit $\rps$ couplings
\begin{eqnarray}
\label{lampterm}
&&\lambda_{333}' (\overline {\nu_{\tau L}^{c}} \ b_L \tilde b^{*}_R
+ \bar b_R^{} b_L^{}\tilde \nu_{\tau L}^{} + \bar b_R^{} \nu_{\tau L}^{}
\tilde b_L^{}\nonumber\\ 
&&- \overline {\tau^{c}_L} \ t_L \tilde b^{*}_R
- \bar b_R^{} t_L^{}\tilde \tau_L^{} - \bar b_R^{} \tau_L^{}\tilde t_L^{})\ 
+ h.c.
\end{eqnarray}
where the subscript for the coupling indicates the third 
generation for all particles involved, 
and  ${\tau^{c}_L}$ is the charge conjugate field.
The above interactions lead to the top-quark decay to
\begin{eqnarray}
\label{2body}
&&t \to b\tilde{\tau}^+ \to b\ \tau^+\schi,\\
\label{3bodyb}
&&t \to  \sbottom \tau^+ \to b \schi\ \tau^+,\\ 
{\rm and}\quad
\label{3bodyt}
&&t \to \tilde{t}_1 \schi \to b \tau^+\ \schi .
\end{eqnarray}
Motivated by SUSY GUT models, we assume that $\tilde{\tau}$ 
is lighter than the top quark and it decays dominantly to 
$\tau\schi$, as indicated in Eq.~(\ref{2body}). $\sbottom$ is 
heavier and the $t\to \sbottom\tau^+$ channel in Eq.~(\ref{3bodyb})
is a three-body decay via a virtual $\sbottom^*$.
We also take $m_{\stop} + m_{\schi} > m_t$ so that the channels of 
Eq.~(\ref{tsusy}) are kinematically suppressed as we only have the 
three-body decay shown in Eq.~(\ref{3bodyt}) via a virtual 
$\stop^*$.

There exist weak indirect bounds on the coupling 
$\lambda_{333}'$. From the $Z\to \tau^+\tau^-$ partial width
via the $\rps$ radiative corrections \cite{yang,batta}, 
a 2$\sigma$ limit $|\lambda^\prime_{333}|<0.58$ was obtained
for $m_{\tilde b}=100$ GeV. Considering the decay rate of
$B \to \tau\nu_\tau X$ yielded a bound of similar 
size \cite{jonathan}. The bounds become weaker for a heavier 
$\tilde b$. To concentrate on this possibly
large coupling, we assume that the other $\rps$ couplings 
involving light fermions are much smaller. The collider
phenomenology then depends on whether $\schi$ in Eqs.~(\ref{2body}) 
to (\ref{3bodyt}) would decay within the detector or
not, as we will discuss in details. 

The rest of the paper is organized as follows:
In Sec. \ref{section2}, we discuss the two-body and three-body 
top decay widths due to $\lambda^\prime_{333}$ and 
obtain the respective branching fractions. In Sec. \ref{section3}, we study
the new signatures coming from these decays and the sensitivity to 
probe $\lambda^\prime_{333}$ at the Tevatron collider. 
We make some concluding remarks in Sec. \ref{section4}. 

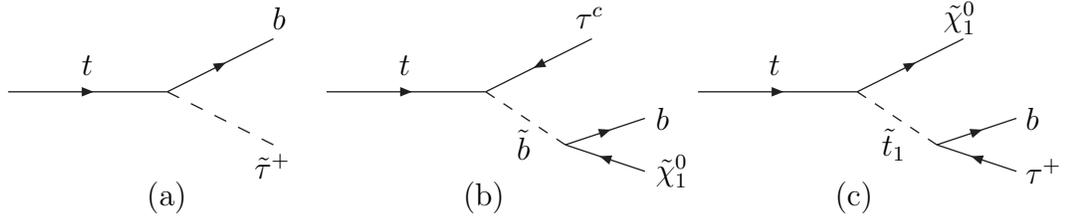
\begin{figure}[tb]
\vskip 36pt
\begin{center}
\centerline{
\begin{picture}(380,80)(-15,-15)
\ArrowLine(0,40)(60,40)
\ArrowLine(60,40)(100,60)
\DashLine(60,40)(100,20){5}
\Text(30,50)[c]{$t$}
\Text(100,68)[l]{$b$}
\Text(100,12)[c]{$\tilde{\tau}^+$}
\Text(60,0)[c]{(a)}
\ArrowLine(120,40)(180,40)
\ArrowLine(220,60)(180,40)
\DashLine(180,40)(210,20){4}
\ArrowLine(210,20)(240,30)
\ArrowLine(240,10)(210,20)
\Text(150,50)[c]{$t$}
\Text(220,68)[c]{$\tau^c$}
\Text(195,20)[c]{$\tilde{b}$}
\Text(245,30)[l]{$b$}
\Text(245,10)[l]{$\schi$}
\Text(180,0)[c]{(b)}
\ArrowLine(260,40)(320,40)
\ArrowLine(320,40)(360,60)
\DashLine(320,40)(350,20){4}
\ArrowLine(350,20)(380,30)
\ArrowLine(380,10)(350,20)
\Text(290,50)[c]{$t$}
\Text(360,68)[c]{$\schi$}
\Text(335,20)[c]{$\tilde{t}_1$}
\Text(385,30)[l]{$b$}
\Text(385,10)[l]{$\tau^+$}
\Text(320,0)[c]{(c)}
\end{picture}
}
\end{center}
\vskip -0.8cm
\caption{Feynman diagrams for top-quark 
two-body (a) and three-body (b,c) decays via 
$R$-parity violation operators in Eq.~(\ref{lampterm}).}
\label{diagrams}
\end{figure}

\section{Top Decays}\label{section2}

We evaluate the partial decay widths for processes  
(\ref{2body}), (\ref{3bodyb}) and (\ref{3bodyt}) depicted 
in Fig.~\ref{diagrams}.
Denoting the two- and three-body decay widths 
by $\Gamma_2$ and $\Gamma_3$, and neglecting the masses of 
$b$ and ${\tau}$, we obtain for the two-body decay 
\be
\label{2width}
\Gamma_2 (t\seta b\tilde{\tau}) = \frac{|\lambda^\prime_{333}|^2}{32\pi} 
m_t \left( 1 - \frac{m^2_{\tilde{\tau}}}{m^2_t} \right)^2 \;, 
\ee
which agrees with Ref.~\cite{roger}; and for the three-body decay
\be
\label{3width}
\Gamma_3 (t\to \tau b \tilde{\chi}^0_1 ) = 
\frac{|\lambda^\prime_{333}|^2}{3072\pi^3}\left( 
(C_L^2 + C_R^2){\frac{m_t^5}{m^4_{\tilde b}}}\; I_{\tilde{b}}(x) + 
(C_L^{'2} + C_R^{'2}){\frac{m_t^5}{m^4_{\tilde t}}}\; 
I_{\tilde{t}}(x)\right)\ ,
\ee
with $C_{L,R}$ and $C^{'}_{L,R}$ being the left- and right-handed coupling 
constants for the vertices $b\,\tilde{b}\,\schi$ and $t\,\tilde{t}\,\schi$
respectively \cite{susy}. 
The phase space integrals $I_i(x)$ are given by 
\be
\label{i}
I_{\tilde{b}}(x) = 12\int^1_{x_{\chi}} dx \frac{(1-x)^2(x - x_{\chi})^2}
{x(1-x/x_{\tilde b})^2}\;,
\ee
and
\be
\label{it}
I_{\tilde{t}}(x) = 12\int^{(1-x_{\chi}^{1/2})^2}_0 dx \frac{x}{
(1-x/x_{\tilde{t}})^2}
\left[(1+x_{\chi}-x)+4x_{\chi}^{1/2}A\right]\lambda^{1/2}(1,x_{\chi},x)\;,
\ee
where $x_{i}= {m^2_i}/{m^2_t}$, 
$A=C^{'}_L C^{'}_R/(C_L^{'2} + C_R^{'2})$
and $\lambda$ is the usual K\"allen kinematical function. 
The factor of 12 is to normalize
the integrals $I_{\tilde{b}}$ and $I_{\tilde{t}}$ to unity in the limit 
$x_{\schi}\to 0$ and 
$x_{\sbottom}$,$x_{\tilde{t}}\gg 1$. The ratio of these two decay widths is 
$\Gamma_2/ \Gamma_3 \approx 48\pi^2 (C_L^2+C_R^2) m^4_{\tilde{t}}/m^4_t$,
which shows that the two-body mode dominates by more 
than three orders of magnitude. 

The SM partial width for the top-quark decay $t\to bW^+$ is
\be
\label{smwidth}
\Gamma_{SM} (t\seta bW^+) = \frac{g^2 m^3_t}{64\pi M_W^2} |V_{tb}|^2 
\left(1 - \frac{m^2_{W}}{m^2_t} \right) \left( 1 + \frac{m^2_{W}}{m^2_t} 
- 2\frac{m^4_{W}}{m^4_t} \right)\;.
\ee
The ratio for the $\rps$ decay width with respect to that of 
the SM is thus 
\begin{equation}
\Gamma_2/\Gamma_{SM}\approx 
2|\lambda^\prime_{333}|^2 M^2_W/g^2m^2_t\approx |\lambda^\prime_{333}|^2.
\end{equation}
We see that a coupling $|\lambda^\prime_{333}|\sim 0.5$ 
would allow a branching ratio of about $20\%$. 

\begin{figure}[htb]
\centerline{
\psfig{file=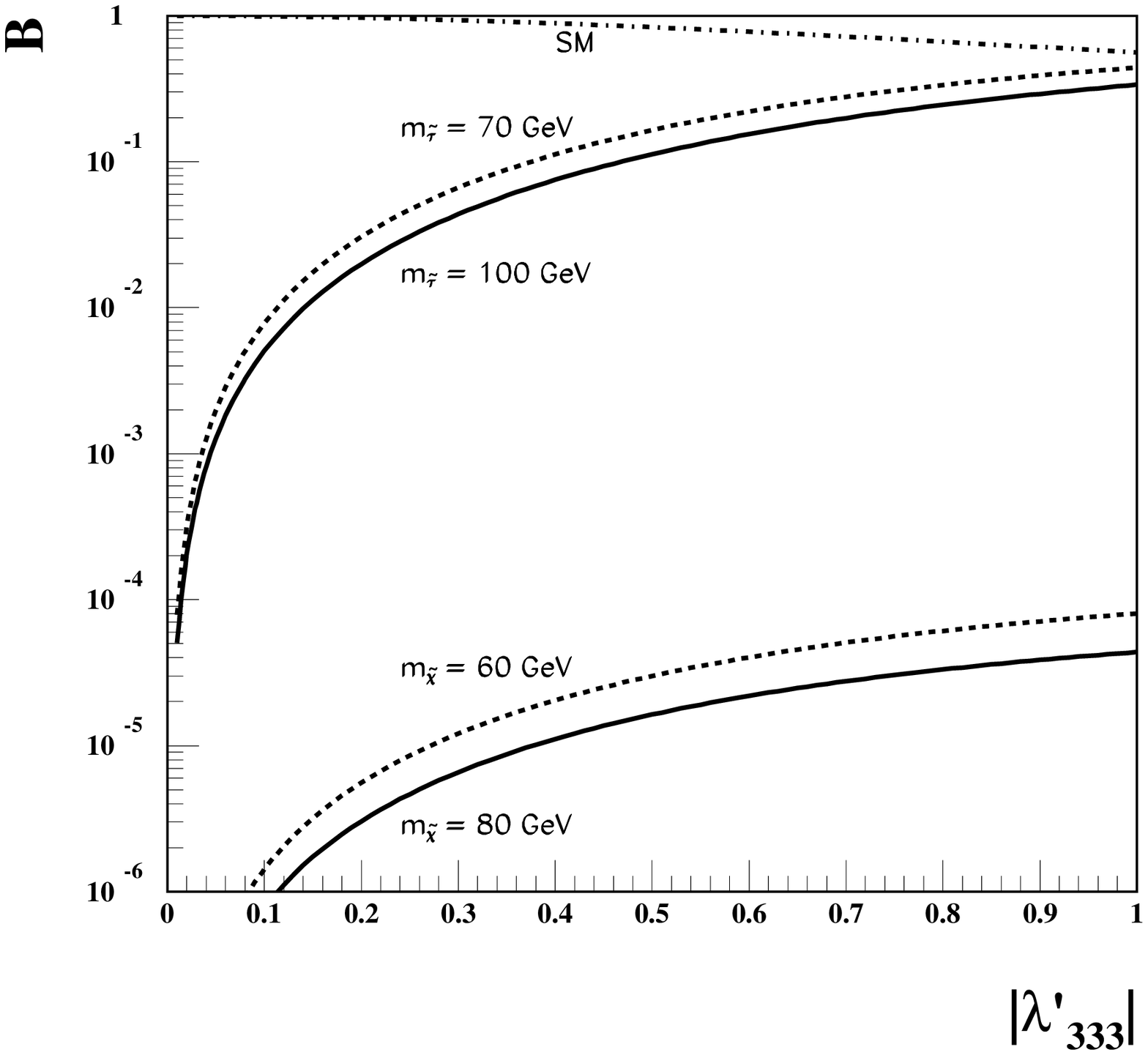,height=9cm,width=11cm}}
\caption[]{$R$-parity violating branching fractions 
for top decays as a function of the coupling constant 
$|\lambda^\prime_{333}|$. The two upper parallel curves 
represent the two-body decay $t \seta b \tilde{\tau}^+$, 
for $m_{\tilde{\tau}}= 70$ GeV (dashed) 
and $m_{\tilde{\tau}} = 100$ GeV (solid)
with $m_{\schi} = 60$ GeV; while the two lower parallel curves 
stand for the three-body decay, 
$t \seta b \tau^+ \tilde{\chi}^0_1$, 
for $m_{\tilde{\chi}^0_1} = 60$ GeV (dashed) and 
$m_{\tilde{\chi}^0_1} = 80$ GeV (solid) with $m_{\tilde{\tau}}= 70$ GeV. 
We assumed $m_{\tilde{b}} = m_{\tilde{t}_1} = 200$ GeV 
for the three-body decay.
The dot-dashed curve at the top is for the SM channel ($t\to W^+b$) 
with $m_{\tilde{\tau}}= 70$ GeV.}  
\label{brfig}
\ef

In Fig.~\ref{brfig} we show the magnitude for the branching fractions 
of the two-body decay for $m_{\tilde{\tau}} = 70$ 
and 100 GeV with $m_{\schi} = 60$ GeV
and three-body decay for $m_{\schi} = 60$ and 80 GeV 
with $m_{\tilde{\tau}} = 70$ GeV, as a function of the 
coupling constant $|\lambda^\prime_{333}|$. 
We have chosen $m_{\tilde{b}} = m_{\tilde{t}_1} = 200$ GeV and 
$(C_L^2 + C_R^2) \approx (C_L^{'2} + C_R^{'2}) \approx 0.1$, with 
$A \ll 0.1$, which enhances the three-body decay and 
is compatible with the 
masses we have assumed for the lightest neutralino. As one can see, the 
two-body decay can be very important reaching $10-20\%$ or higher 
for large $|\lambda^\prime_{333}|$ and therefore could play an 
important role in the top decay phenomenology. On the other hand, the 
three-body decay is too small to be significant. For illustration,
we have also included the SM channel ($t\to W^+b$) 
in Fig.~\ref{brfig} at the top with $m_{\tilde{\tau}}= 70$ GeV.

\begin{figure}[htb]
\centerline{
\psfig{file=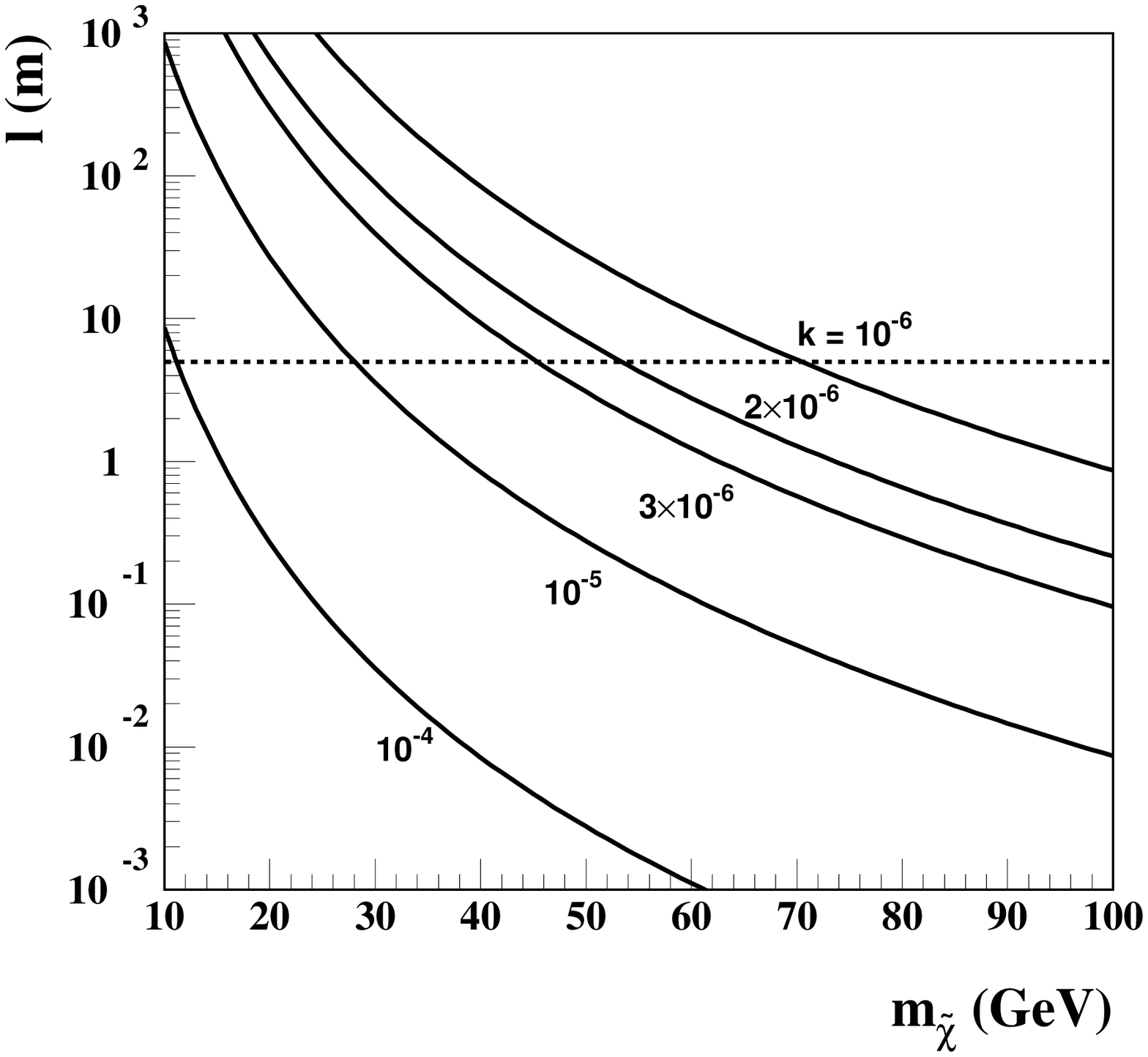,height=9cm,width=11cm}}
\caption[]{Estimated decay length for $\schi\to \nu_\tau b\bar b$ 
with various effective coupling $\kappa$, assuming $\gamma\beta=3$.
The dashed line indicates the five-meter mark.
}
\label{decayL}
\ef

\section{Signatures}\label{section3}

We base our analyses on the SM top-quark production at the Tevatron
\begin{equation}
p\bar p \to t\bar t X,
\end{equation}
from those recently reported in the CDF/D0 experiments \cite{top}.
The $t\bar t$ signal can be identified above the SM backgrounds
from their distinctive decay products. 
In our analyses, the event acceptance can be expressed by
\be
{\cal A}=\sum_{i,j}^{SM,\rps} \epsilon_{ij}\ B_i\ B_j,
\ee
where $B_{SM}$ and $B_{\rps}$ are the branching fractions
of $t$ and $\bar t$ decays to $t\to bW$ and $t\to b\tau\schi$,
$\epsilon_{ij}$ the corresponding detection efficiencies.

With the $\rps$ interactions under consideration, the LSP $\schi$ 
is not stable and will decay via $\schi\to \nu_\tau b\bar b$. 
The collider phenomenology will then depend upon if it
decays within the detector or not. 
The decay length is calculated by $\ell = \gamma\beta\ c\tau$, where
the $\schi$ life-time $\tau$ can be written as \cite{rplimit}
\begin{equation}
\tau^{-1} = \kappa^2 {{ {3\, G_F^2 m^5_{\schi}} \over {192\pi^3}} },
\end{equation}
where the effective coupling $\kappa$ depends on the composition 
of the LSP and the masses of the virtual 
sfermions \cite{rplimit,agashe}, and is typically
$\kappa \sim 0.1\ (100\ {\gev}/m_{\tilde{f}} )^2\ \lambda^\prime_{333}$.
At the Tevatron energies, the $\gamma\beta$ factor for $\schi$ is
typically about $1\sim 5$. With this estimation, we can calculate
the $\schi$ decay length as shown in Fig.~\ref{decayL}.
We see that for a rather light LSP
$m_{\schi}\sim {\cal O}$(10 GeV) or a weak effective coupling
$\kappa<10^{-5}$, $\schi$ may escape from detection before
it decays, resulting in missing energy as the standard SUSY 
signal.
In fact, there is no mass constraint of this order for the 
LSP for general SUSY models. Also, such a small coupling is
possible even with $\lambda^\prime_{333}\sim {\cal O}(0.1 -1)$ 
if the third generation sfermions are very massive about 10 TeV
as advocated in certain ``inverse hierarchy'' models \cite{invH}. 
On the other hand, for a large region of the parameter space in
$(\kappa, m_{\schi})$,  $\schi$ will decay within the detector
of a few meters. We will therefore study both cases below.

\subsection{ Long-lived $\schi$ as missing transverse energy}

\begin{figure}[htb]
\centerline{
\psfig{file=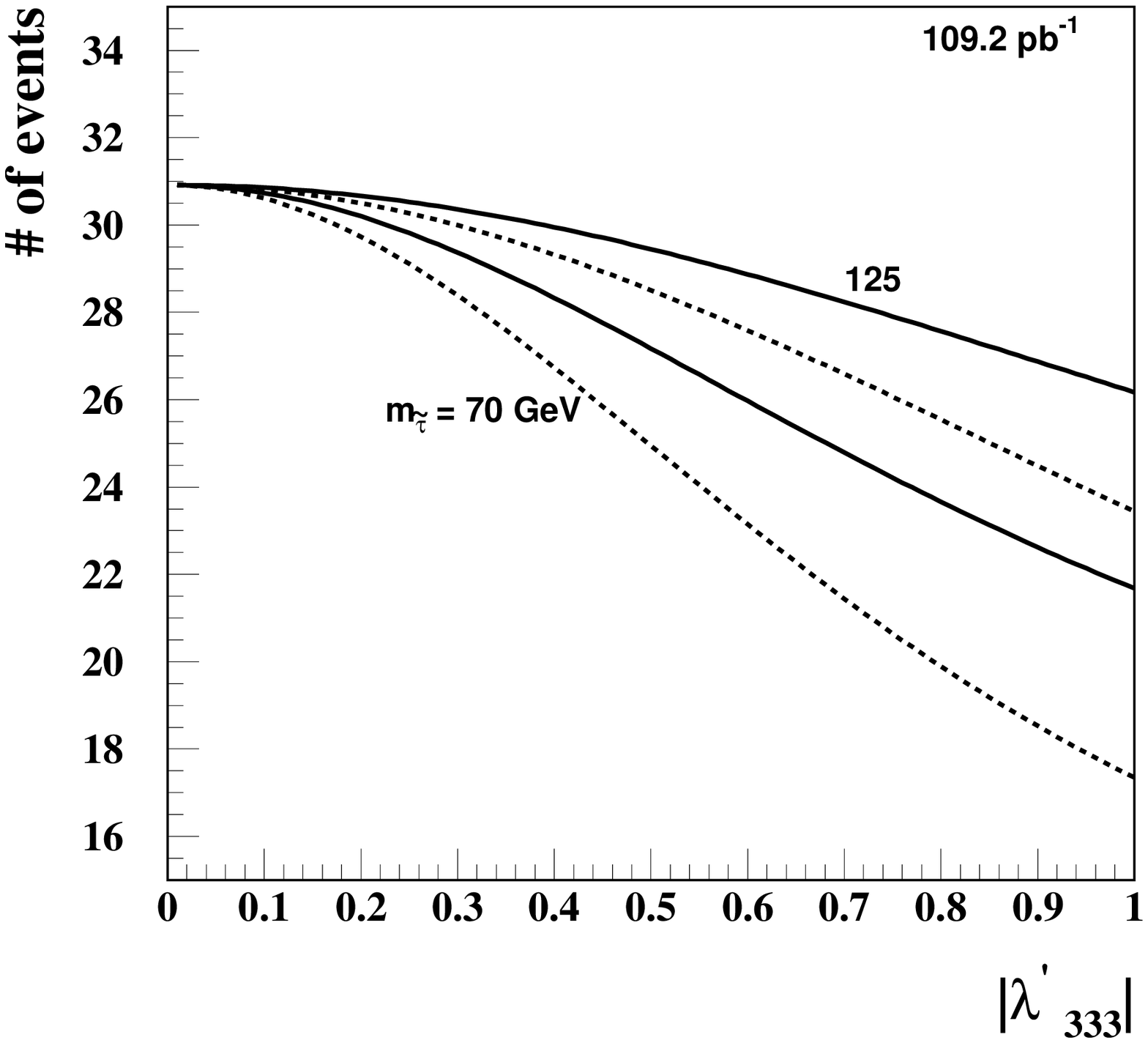,height=9cm,width=11cm}}
\caption[]{Predicted number of events versus the coupling
$|\lambda^\prime_{333}|$ for Run I ($\sqrt s=1.8$ TeV) with the 
D0 event reconstruction method and a data sample of 109.2 pb$^{-1}$
for two representative values of $m_{\tilde{\tau}}$. The solid 
curves stand for a long-lived $\schi$ and the dashed ones for 
$\schi$ to decay within detector.}
\label{nev}
\ef

We consider the simple situation first: The LSP $\schi$ is
stable in the experimental scale, resulting in missing 
transverse energy. It is important to note that in this 
case the top-quark decay via the
$\rps$ coupling $\lambda^\prime_{333}$ yields the same signal 
as for the charged Higgs boson search $H^+\to \tau^+\nu b$. 
The efficiencies for these final states are given by \cite{d0}
\be
\begin{array}{ccc}
\epsilon_{ij}\   ($\%$)      & t \seta bW & t \seta  b\tau\nu  \\
t \seta bW      & 3.42       & 1.36               \\
t\seta b\tau\nu & 1.36       & 0.41    
\label{tab}
\end{array}
\ee
We adopt the D0 analyses and assume the same efficiencies
for the event reconstruction. The predicted event rate can
thus be obtained by
\be
N = {\cal A}\ \sigma(t\bar t)\ {\cal L} + N_B,
\label{Nevents}
\ee
where $\sigma(t\bar t)\approx 5.5$ pb \cite{d0}, ${\cal L}$ is the
integrated luminosity. $N_B$ presents the expected non-top-quark
SM backgrounds, which includes $W+$jets events and QCD multi-jet events with 
a misidentified lepton and large $\met$ \cite{d0}.
Based on this event reconstruction
scheme, we calculate the predicted $t\bar t$ signal events
as shown by the solid curves in Fig.~\ref{nev}. 
We have included all the four
channels listed in Eq.~(\ref{tab}). 
The SM expectation for $t\bar t$ events is implied
at $|\lambda^\prime_{333}|=0$. 
The larger the $\rps$ coupling is, the smaller the SM branching
fraction ($B_{SM}$) would be, as seen from the top curve
in Fig.~\ref{brfig}. Since the detection efficiencies are
optimized for the SM decay channels, the total
predicted number of $t\bar t$ events decreases for higher values
of $|\lambda^\prime_{333}|$ after convoluting with
the efficiencies of Eq.~(\ref{tab}), resulting in
a ``disappearance'' experiment.

\begin{figure}
\begin{center}
\begin{minipage}[t]{130mm}  
{\setlength{\unitlength}{1mm}
\begin{picture}(60,140)
\put(8,80){\mbox{\psfig{figure=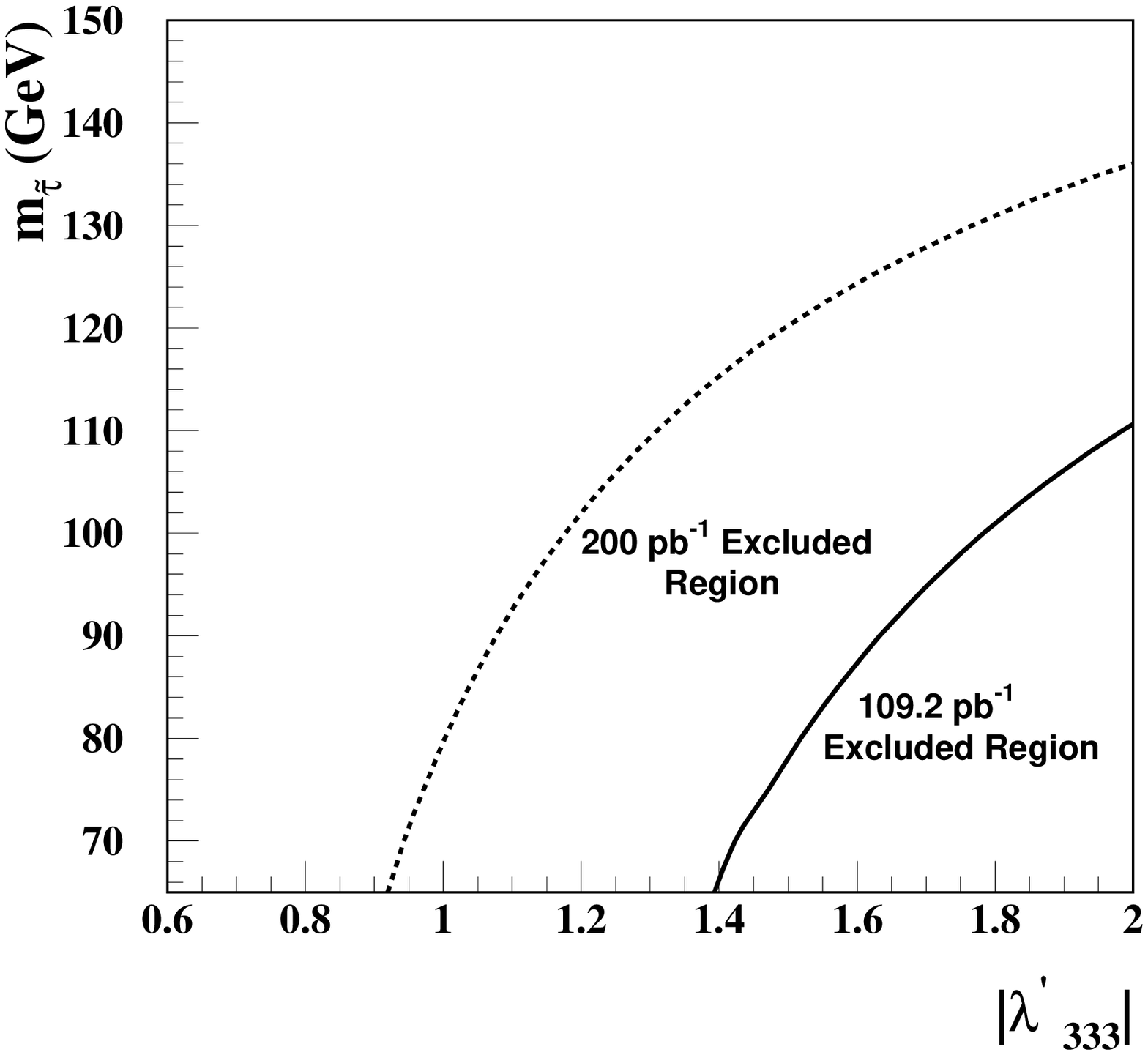,height=5cm}}}
\put(8,-5){\mbox{\psfig{figure=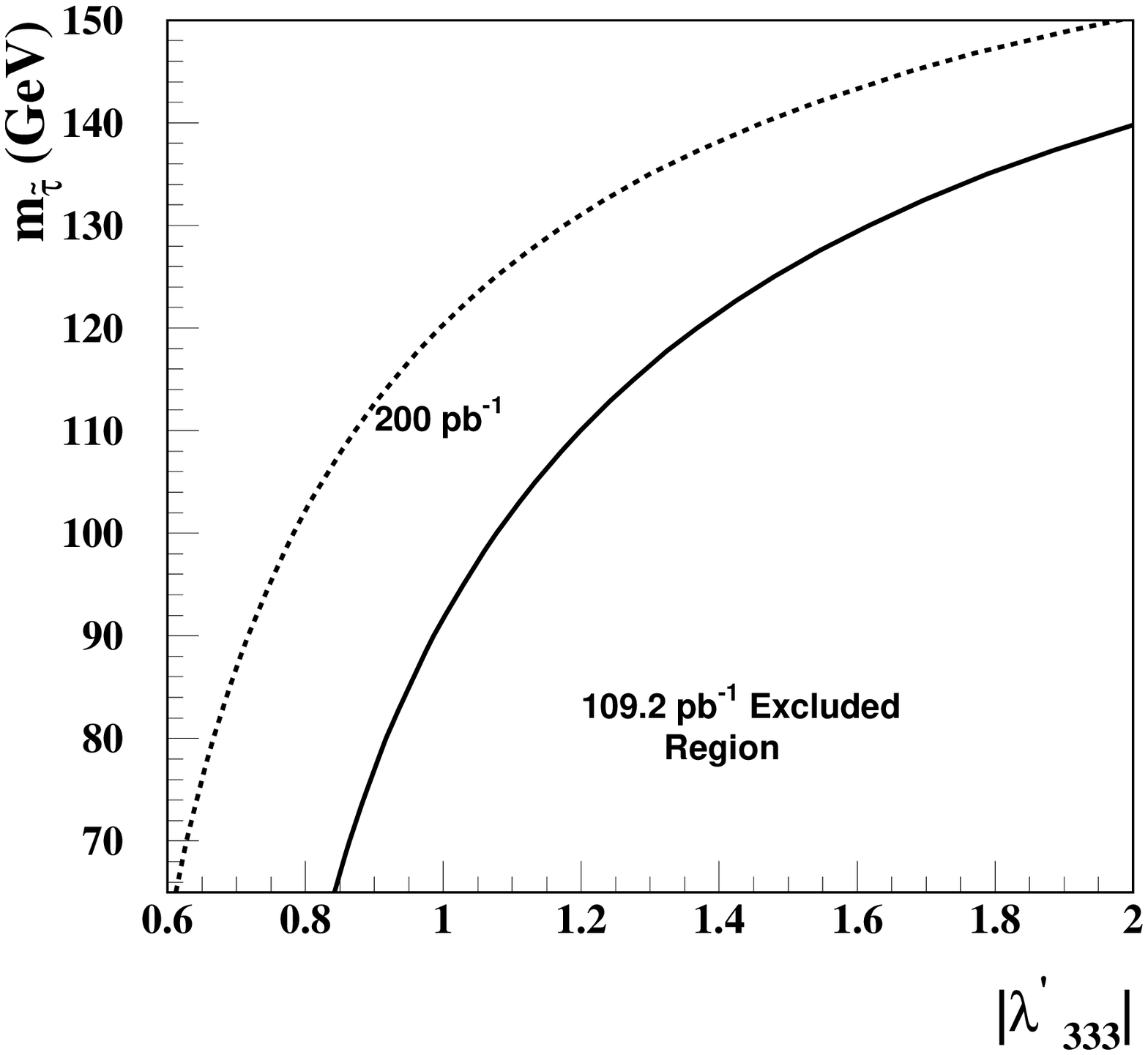,height=5cm}}}
\put(32,165){\makebox(0,0)[bc]{{\small (a)}}}
\put(32,80){\makebox(0,0)[bc]{{\small (b)}}}
\end{picture}}
\end{minipage}
\end{center}
\caption{Contour plots for Run I
in the $m_{\tilde \tau}-|\lambda^\prime_{333}|$ 
plane for exclusion at the $95\%$ confidence level for the cases where 
(a) the $\schi$ is long-lived and (b) the $\schi$ decays within the 
detector. The solid curves are for the data sample of 109.2 pb$^{-1}$ (D0),
and the dashed curves are for 200 pb$^{-1}$.}
\label{tev}
\ef

The non-observation of the signal beyond the SM 
at the Tevatron experiments can be translated to constraints 
on $m_{\tilde \tau}$ and $|\lambda^\prime_{333}|$.
In Fig.~\ref{tev}(a), we show the exclusion limit in the
$m_{\tilde \tau}-|\lambda^\prime_{333}|$ plane at the 95\%
confidence level. Here we have adopted the Poisson
statistics treatment because the number of signal events
with the existing data is not expected to be very large.
Since our procedure closely follows the
D0 analysis \cite{d0}, we assume the D0 data sample 
of 109.2 pb$^{-1}$ and obtain the solid curve. The dashed curve 
is for 200 pb$^{-1}$ (CDF and D0). The regions below 
the curves are excluded at a 95\% C.L.~level. 
For the current mass limit $m_{\tilde \tau}\approx 70$ GeV \cite{stau},
we see that the existing Tevatron
data can put a bound on the coupling to be 
$|\lambda^\prime_{333}|<0.94$.
The bound becomes weaker for a heavier $m_{\tilde \tau}$.
At Run II with $\sqrt s=2$ TeV and 
$\sigma(t\bar{t}) \approx 6.8$ pb, the sensitivity
is expected to be greatly enhanced. This is shown in
Fig.~\ref{run2}(a). For $m_{\tilde \tau}=70$ GeV, the $2\sigma$
bound is $|\lambda^\prime_{333}|<0.38\ (0.24)$ 
with 2 (10) fb$^{-1}$. 
This bound is essentially independent of the other SUSY
parameters, and is more stringent than the indirect one 
obtained in Ref.~\cite{yang,batta}. For comparison, the 3$\sigma$
contours are also given by the dashed curves in the figure.

\begin{figure}
\begin{center}
\begin{minipage}[t]{130mm}  
{\setlength{\unitlength}{1mm}
\begin{picture}(60,140)
\put(8,80){\mbox{\psfig{figure=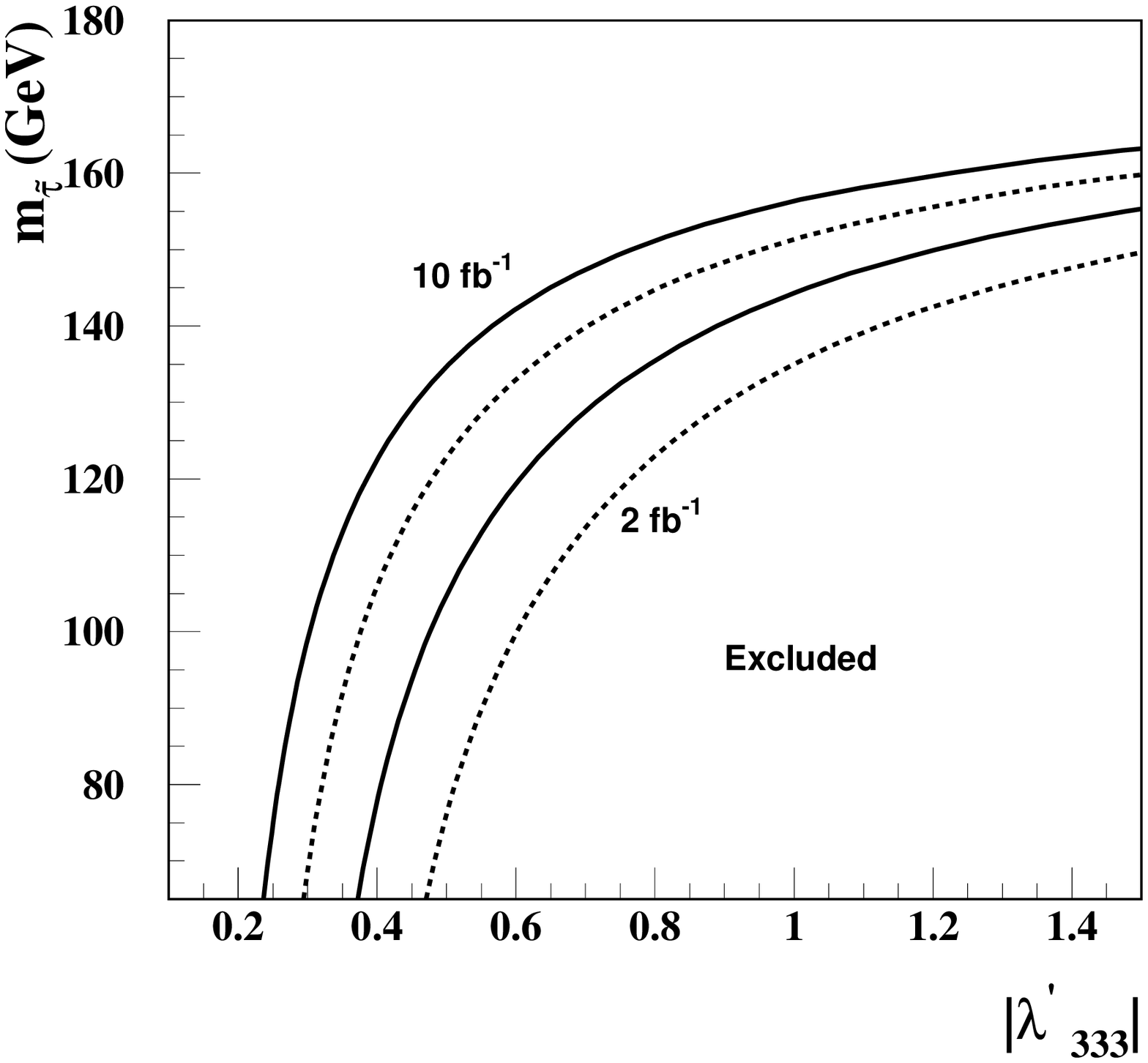,height=5cm}}}
\put(8,-5){\mbox{\psfig{figure=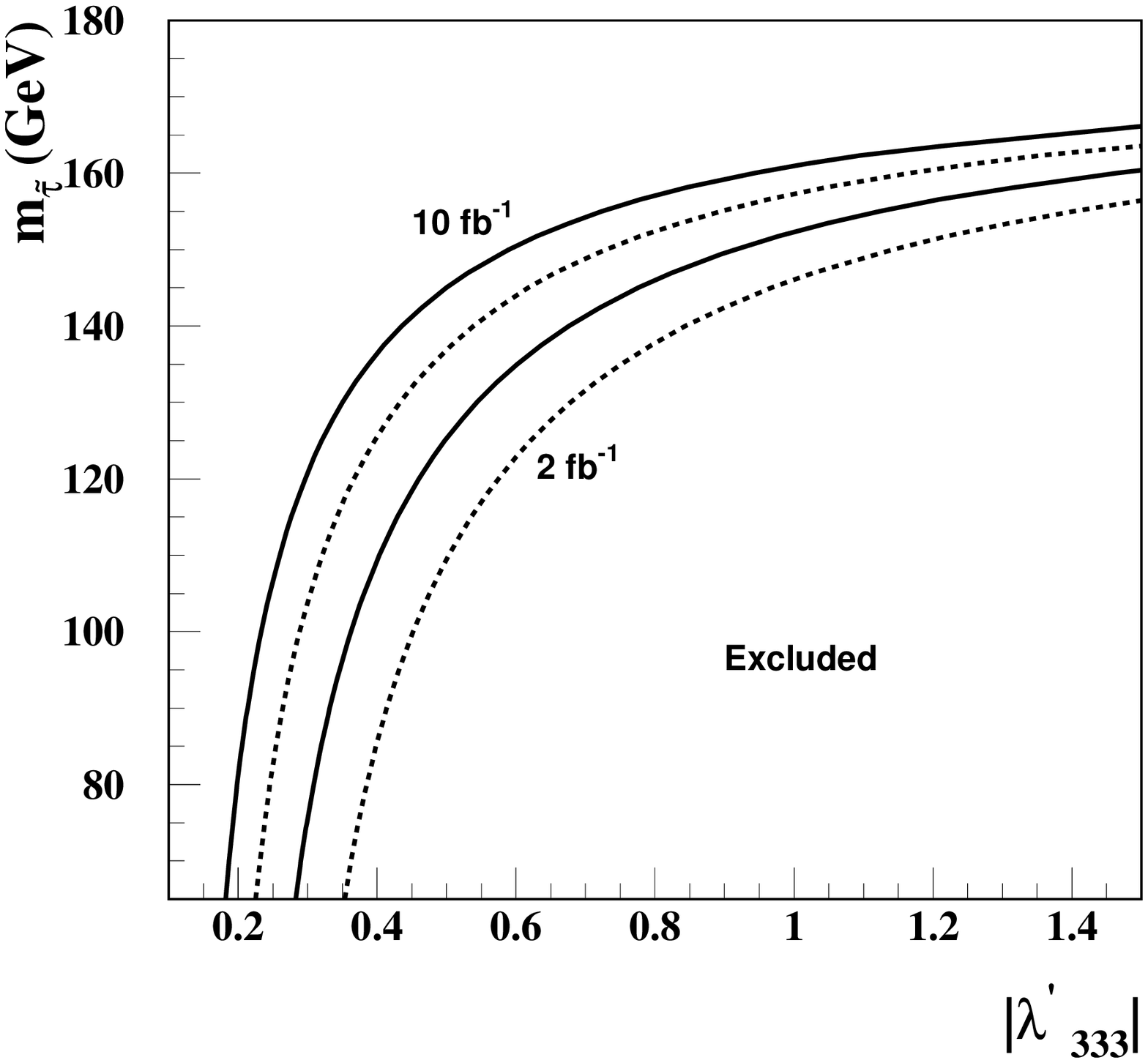,height=5cm}}}
\put(32,165){\makebox(0,0)[bc]{{\small (a)}}}
\put(32,80){\makebox(0,0)[bc]{{\small (b)}}}
\end{picture}}
\end{minipage}
\end{center}
\caption{Contour plots in the $m_{\tilde \tau}-|\lambda^\prime_{333}|$ 
plane for exclusion at the $2\sigma$ (solid) and $3\sigma$ (dashed) limits 
at Run II with 2 and 10 fb$^{-1}$,
for the cases where (a) the $\schi$ is long-lived and 
(b) the $\schi$ decays within the detector.}
\label{run2}
\ef

\subsection{$\schi\to \nu_\tau b\bar b$}

As seen in Fig.~\ref{decayL}, if $m_{\schi}$ and $\kappa$ are large
$\schi$ can decay within detector, yielding extra $b\bar b$ plus 
missing transverse energy in the final state.
Two more $b$ quarks would make the signal
very different from the SM mode. This decay was considered
at the Tevatron in the light of the total $t\bar t$ production
cross section and a rather loose bound was 
inferred \cite{agashe,jonathan}.
We here consider two searching scenarios.

First, we still adopt the disappearance method because the study
of SM $t\bar t$ events is well established. Since the signal mode
\begin{equation}
t \to \tau^+ b \schi \to \tau^+ b\ b\bar b \nu_\tau
\label{dec}
\end{equation}
is significantly different from the mode 
$t \to \tau b \nu_\tau$ being searched for,
the signal events can be effectively removed from the sample,
leading to a lower acceptance efficiency. 
We thus keep the SM $t\to bW$ mode with the same
efficiency as in Eq.~(\ref{tab}), while take the others 
to be zero. This is an optimistic choice for the signal 
searches in a disappearance experiment, but it will not 
significantly alter the results. 
The ``more disappearance'' results in a larger
deviation from the SM expectation of $t\bar t$ production,
as shown by the dashed curves in Fig.~\ref{nev}. This in turn
leads to more stringent constraints with the existing Tevatron
data as shown in Fig.~\ref{tev}(b) and 
with Run II data in Fig.~\ref{run2}(b).
The current Tevatron data can put a bound on the coupling 
to be about $|\lambda^\prime_{333}|<0.63$ for the short-lived
$\schi$ with $m_{\tilde \tau}\sim 70$ GeV.
At Run II, the $2\sigma$ bound will be 
$|\lambda^\prime_{333}|<0.29\ (0.19)$ 
with 2 (10) fb$^{-1}$. 

Alternatively, we could consider the direct search
for the signal channel Eq.~(\ref{dec}). To have a
clean signal, one leptonic decay of 
$t\to b \ell\nu\ (\ell=e,\mu)$ is needed in combination
with the signal mode. This $\ell \tau$ channel 
has been recently studied
experimentally in Ref.~\cite{cdftau}. Unfortunately,
the low branching fraction for the leptonic modes
and the low detection efficiency for the $\tau$ lepton
render the signal events too low to be appreciable.
Only with a luminosity significantly higher than
the Run II projection could this search become 
possibly useful. We note that direct production
processes $p\bar p \to t\tilde\tau X$ have been
suggested to search for this $\rps$ interaction \cite{FB}
and it is worthwhile exploring further.

\section{Discussions and conclusion}\label{section4}

Our analyses closely follow the study reported in Ref.~\cite{d0}
for $H^\pm \to \tau^\pm \nu_\tau$; while our signal for a long-lived
$\schi$ is
for ${\tilde\tau}^\pm \to \tau^\pm \schi$. Although one may expect
to have a softer $\tau$ lepton in our signal than that 
in Ref.~\cite{d0} due to a more
massive $\schi$, we would like to
argue that the event reconstruction efficiency may be quite
comparable since the signal under consideration carries
larger missing energy, which should help in event selection.

A more interesting question to ask would be that if there is a
$\tau^\pm+$missing energy signal established beyond the SM 
expectation, could we tell if it is from a $H^\pm$ decay or from 
${\tilde\tau}^\pm$? To properly address this question, one
should realize that the $\rps$ interactions in Eq.~(\ref{lampterm})
violate lepton number. Therefore, $H^\pm$ and ${\tilde\tau}^\pm$
cannot be distinguishable in terms of the SM quantum numbers.
One would need to specify the complete theory and work out the
allowed state mixing \cite{mix}. In this sense, the Higgs fields lose 
the identity as that in $R$-conserving SUSY theories.

In conclusion, we have studied the top-quark decays 
$t\to \tilde \tau b$ and $t\to \tau b \schi$
via $R$-parity violating interactions in SUSY theories. 
We found that the existing Tevatron data can put a $95\%$
confidence level upper bound on the coupling to be
0.94 (0.63) for a long-lived (short -lived) $\schi$ 
with $m_{\tilde\tau}=70$ GeV. At Tevatron Run II with an 
integrated luminosity of 2 (10) fb$^{-1}$, one can obtain 
a 2$\sigma$ constraint as 0.38 (0.24) for a long-lived $\schi$ 
and 0.29 (0.19) for a short-lived $\schi$,  beyond the current
indirect limit. These limits from the direct search are largely
independent of other SUSY parameters.

\noindent
{\it Note added:} After the submission of this paper, we received
an article \cite{tatsu} in which hadronic observables of $Z$ decays
were studied and new constraint on $|\lambda^\prime_{333}|$ 
was obtained.


\acknowledgments
We thank J. Feng, S. Mrenna, W. Porod and J.-M. Yang
for discussions on the related topics. 
This work was supported in part by a DOE grant
No. DE-FG02-95ER40896, in part by the Wisconsin Alumni Research 
Foundation and in part by Funda\c{c}\~ao de Amparo \`a Pesquisa 
do Estado de S\~ao Paulo (FAPESP).


\end{document}